# Mid-infrared spontaneous and stimulated emission dynamics in black phosphorus

*Hisashi Sumikura,*[*,†,‡] *Akihiko Shinya,*[†,‡] *and Masaya Notomi*[†,‡]

[†]Basic Research Laboratories, NTT, Inc., Atsugi, Kanagawa 243-0198, Japan

[‡]NTT Nanophotonics Center, NTT, Inc., Atsugi, Kanagawa 243-0198, Japan

**ABSTRACT:**

Black phosphorus (BP) has attracted attention as a light-emitting layered semiconductor for mid-infrared (MIR) photonics owing to its direct and thickness-tunable bandgap energy, highly anisotropic optical transitions, and potentially suppressed Auger recombination. However, spontaneous and stimulated emission dynamics governed by carrier recombination in BP have remained unexplored because time-resolved emission spectroscopy is challenging in the MIR region. Here we develop a time-resolved MIR emission microscope combining wavelength upconversion with superconducting single photon detection. This approach enables observation of emission dynamics in photoexcited BP at a wavelength around 4.6 μm with sub-100-ps temporal resolution. Temperature-dependent measurements reveal a crossover from excitonic to electron-hole plasma emission at around 70 K, supported by independent results in the characteristic transition temperature, pump-fluence dependence, and rise and decay dynamics of MIR emission. In a suspended BP structure, we further observe a nonlinear increase in the emission intensity above a well-defined pump threshold. Spectral narrowing and intense picosecond emission pulses appear above the threshold, providing solid evidence of stimulated emission assisted by optical feedback from a Fabry-Pérot cavity. These results elucidate ultrafast carrier dynamics and optical gain formation in BP and establish time-resolved upconversion spectroscopy as a powerful approach for investigating MIR photonic materials and devices.

Black phosphorus (BP), an allotrope of low-cost and widely available phosphorus, has attracted attention as a light-emitting layered semiconductor for mid-infrared (MIR) photonics. Its direct and thickness-tunable bandgap spans a near- to mid-infrared spectral range while maintaining strong photon emission from the monolayer to the bulk limit [1-3], in contrast to transition-metal dichalcogenides, whose direct bandgap and strong emission exist only in the monolayer [4, 5]. BP exhibits anisotropic optical transitions with the large dipole moment aligned along the armchair crystal direction [6, 7]. This anisotropy is advantageous to achieve efficient optical coupling between BP and photonic structures such as waveguides and optical cavities. Furthermore, the nearly identical electron and hole effective masses originating from its symmetric band structure at the Z point are expected to suppress nonradiative Auger recombination at high carrier densities [8, 9]. This characteristic is particularly attractive for fabricating MIR light sources because commonly used narrow-bandgap III-V semiconductors suffer from the severe Auger loss of excited carriers [10-12]. Although BP is susceptible to ambient air owing to its high chemical reactivity, recent advances in surface passivation have improved its environmental stability and expanded its device potential [13-15].

Recent studies have demonstrated BP-based MIR light-emitting devices and optically pumped MIR lasers using distributed-Bragg-reflector cavities [16-20]. However, these studies have relied on steady-state measurements of the emission intensity and spectra, leaving the emission dynamics unexplored. The fundamental questions remain regarding the mechanisms determining the emission lifetime, the evolution of photon emitting states with temperature, and the temporal dynamics of stimulated emission. Addressing these issues requires time-resolved MIR emission spectroscopy, which is still challenging because ultrafast and highly sensitive MIR photon detectors are limited.

Here, we develop a time-resolved MIR emission microscope based on wavelength upconversion of MIR photons in a nonlinear optical crystal and superconducting single photon detection, enabling observation of ultrafast MIR emission dynamics with sub-100-ps temporal resolution. Using this technique together with time-integrated spectroscopy, we investigate the polarization, excitation, and temperature dependences of spontaneous and stimulated emission in photoexcited BP. We reveal the impact of nonradiative surface states on the emission lifetime, identify a crossover from excitonic to electron-hole plasma (EHP) emission near 70 K, and observe intense picosecond stimulated emission assisted by a Fabry-Pérot cavity formed in a suspended BP structure.

## Spontaneous emission in bulk BP

To investigate spontaneous emission dynamics in bulk BP, we first measured the time-integrated MIR emission intensity for a thick region of a BP flake (Fig. 1a) as a function of excitation and detection polarizations. Figure 1b shows the polarization dependence of the emission intensity. The intensity reaches a maximum when both the excitation and detection polarizations are aligned near 55°, which indicates strong anisotropy in both optical absorption and emission. This result is consistent with the anisotropic transition dipole moment of BP oriented along the armchair crystal direction [6, 7].

Under the optical excitation polarized parallel to the armchair direction (50°), we next measured the emission intensity at four different positions on the BP flake as a function of pump fluence. As shown in Fig. 1c, although the absolute emission intensity varies across the BP flake, all positions exhibit a nearly linear dependence on the pump fluence below 0.4 mJ $cm^{-2}$ per pulse. Above this pump fluence, the emission intensity begins to saturate slightly before increasing again at the pump fluences exceeding 4 mJ $cm^{-2}$ per pulse. At the highest pump fluence of 8.6 mJ $cm^{-2}$ per pulse, a narrow spectral peak appears at a wavelength of 4.6 μm, whereas the emission spectrum remains broad at a lower pump fluence (Fig. 1d).

Figure 1e shows the time-resolved emission measured at the same four positions at 0.37 mJ $cm^{-2}$ per pulse, where the intensity saturation is negligible. The peak counts are almost identical for all positions, indicating comparable initial populations of photoexcited emitters. In contrast, the decay dynamics exhibit spatial variation. The emission transients are well expressed by a single-exponential decay with lifetimes ranging from 0.48 to 0.78 ns. The spatial variation in the emission decay is correlated with the integrated emission intensity shown in Fig. 1c. As seen in Fig. 1f, the emission intensity decreases as the emission decay rate increases. The inverse relationship is clearly observed for all positions.

To further examine the dominant mechanism determining the emission decay, the BP flake was exposed to ambient air for 30 minutes. After the air exposure, the emission intensity decreases while the decay rate increases at positions labeled by #1 and #2 (Fig. 1f). Only minor changes are observed at positions #3 and #4. This trend suggests that the spatial variation in the emission intensity is governed by differences in nonradiative recombination rather than by changes in spontaneous emission recombination and the initial emitter population. Since air exposure oxidizes BP surfaces [13, 14], the generated surface states enhance nonradiative recombination. The small responses at positions #3 and #4 suggest that nonradiative recombination in these regions is already dominated by pre-existing surface states or other defects, which restricts the additional impact of surface oxidation.

The inverse relationship between the emission intensity $I$ and the observed decay rate $\Gamma$ can be explained by a simple relation. The emission intensity is given by $I = \eta I_0$, where $\eta$ is the internal quantum efficiency and $I_0$ is the emission intensity at $\eta = 1$. Using $\eta = \Gamma_R/(\Gamma_R + \Gamma_{NR}) = \Gamma_R/\Gamma$, where $\Gamma_R$ and $\Gamma_{NR}$ are the radiative and nonradiative decay rates, respectively, yields $I = I_0\Gamma_R/\Gamma$. Since the intrinsic $I_0$ and $\Gamma_R$ are expected to be constant across the measured positions, the inverse relationship observed in Fig. 1f indicates that the decay dynamics are primarily governed by the nonradiative decay rate $\Gamma_{NR}$.

**Exciton to electron–hole plasma crossover**

To identify the origin of spontaneous emission, we performed time-integrated and time-resolved measurements as the sample temperature changes. Figure 2a shows the emission intensity as a function of pump fluence at different temperatures. The overall emission intensity decreases with increasing temperature, while saturation at high pump fluences becomes more pronounced. The emission intensity $I$ follows the power-law dependence on the pump fluence $P$, described by $I = \alpha P^k$, where $\alpha$ and $k$ are a temperature-dependent factor and a power-law exponent, respectively. Figure 2b shows the temperature dependence of $k$, obtained by fitting the low-fluence data of the emission intensity. At low temperatures below 40 K, the emission intensity exhibits a weak but reproducible superlinear dependence on the pump fluence with $k \approx 1.15$. Above 70 K, the power-law exponent decreases with increasing temperature and reaches $k = 1.00$ at room temperature.

Figure 2c shows the corresponding emission dynamics measured at a pump fluence of 0.37 mJ $cm^{-2}$ per pulse. Below 40 K, the emission transients exhibit a single-exponential decay, and their peak counts remain almost constant in this low temperature range. The emission peak is delayed by approximately 0.1 ns at 4 K compared with 250 K. The emission dynamics change qualitatively above 70 K. The emission rise becomes faster, and the peak count decreases. The emission decay exhibits a non-exponential behavior accompanied by an additional slow component in the late time of the decay.

To characterize this behavior quantitatively, the emission transients were fitted using single- or double-exponential functions. Figure 2d shows the extracted decay rates as a function of temperature. The fast decay component remains approximately 1.5 $ns^{-1}$ in the measured temperature range. In contrast, an additional slower component appears above 70 K and decreases from 0.9 $ns^{-1}$ to 0.07 $ns^{-1}$ as the temperature increases.

These experimental results suggest a crossover in the dominant emissive state from excitonic emission below 40 K to electron–hole plasma (EHP) emission above 70 K, which is independently supported by following interpretations. First, the observed transition temperature (~70 K) is comparable to the thermal dissociation temperature $T_e$ of excitons estimated from the reported exciton binding energy of bulk BP ($E_b$ = 9.0 meV) [21, 22]. Using the relation by $E_b = k_B T_e$, where $k_B$ is the Boltzmann constant, yields $T_e \approx 100$ K. Second, the low-temperature emission exhibits a reproducible superlinear pump-fluence dependence with a power-law exponent of $k \approx 1.15$ (Fig. 2b). This superlinear behavior is a characteristic of free- or bound-exciton recombination, reflecting the two-particle process of exciton formation from photoexcited electrons and holes [21, 23]. Third, the time-resolved measurements reveal a delayed rise of the emission transient at low temperatures (Figs. 1e and 2c), indicating the formation of cooled excitons following photoexcitation of electrons and holes [24-26]. Finally, the emission dynamics change qualitatively above approximately 70 K, where an additional slow decay component develops (Figs. 2c, d). Such behavior cannot be explained by the introduction of an additional one-particle nonradiative recombination channel. The nonradiative process contributes linearly to the carrier-loss rate and therefore modifies the overall exponential decay rate but does not generate a distinct slow decay component.

The theoretical support for the last interpretation is provided by the rate equation. In EHP, the photoexcited carrier density $N(t)$ follows

$$dN/dt = - AN - BN^2 - CN^3,$$

where $A$, $B$, and $C$ are the nonradiative Shockley-Read-Hall (SRH) recombination rate, two-particle (bimolecular) radiative recombination coefficient, and nonradiative Auger recombination coefficient, respectively. Under the present excitation conditions, Auger recombination is negligible because the pump fluence is below the onset of emission saturation. The SRH recombination term $AN$ originating an exponential decay of $N$ is a linear combination of one-particle nonradiative processes. Thus, the decay rate $A$ means a sum of all decay rates of nonradiative processes regardless of whether the decay is fast or slow. The radiative term $BN^2$ gives rise to a non-exponential decay whose rate depends on the photoexcited carrier density, which can be observed as the slow component at elevated temperatures. The photoexcited carrier density at the onset of measured emission dynamics is estimated to be $N(0) = 1.7 \times 10^{18}$ cm$^{-3}$ from the pump fluence and the penetration depth of pump light. Fitting the measured transient with the rate equation yields a radiative recombination coefficient of $B = (4.7 \pm 0.3) \times 10^{-10}$ cm$^3$ s$^{-1}$ at 200 K. (see Supplementary) This estimated value agrees well with previously reported values of (2-6) × $10^{-10}$ $cm^3$ $s^{-1}$ [27-29]. In contrast, excitonic recombination can be described by a one-quasiparticle recombination model in which the exciton density decays exponentially through both radiative and nonradiative processes, resulting in the single-exponential emission decay observed below 40 K. Nonradiative two-quasiparticle recombination processes including exciton-exciton interaction are negligible because the pump fluence is below the onset of emission saturation.

To summarize, the characteristic transition temperature, pump-fluence dependence, and rise and decay dynamics provide independent and mutually consistent signatures of an excitonic to EHP emission crossover in bulk BP. Notably, the instantaneous initial areal carrier density of 1.5 × $10^{13}$ $cm^{-2}$ exceeds the nominal Mott critical density reported for BP (~$10^{12}$ $cm^{-2}$) by one order of magnitude [30]. Nevertheless, the low-temperature emission remains well described by a single-exponential decay, suggesting that excitonic emission persists throughout the observed recombination dynamics despite the high initial carrier density.

**Stimulated emission in a suspended BP structure**

To demonstrate stimulated emission assisted by an optical cavity, we measured a suspended BP structure naturally formed near the edge of the BP flake (Fig. 1a and Supplementary). The suspended structure creates a gap between the BP film and the underlying $SiO_2$/Si substrate, forming a vertical Fabry-Pérot cavity that provides optical feedback for the emitted MIR photons. Figure 3a shows the spatial distribution of the spectrally integrated emission intensity measured at 4 K. Periodic interference fringes are clearly observed throughout the suspended region, indicating spatial modulation of the optical cavity resonance caused by slight bending of the suspended BP film and corresponding gradual variations in the gap thickness. The cavity formation is further confirmed by the MIR reflection spectrum, which exhibits periodical peaks from the Fabry-Pérot resonances. From the measured spectral period, the cavity length is estimated to be approximately 12 μm, which is in excellent agreement with the height of the suspended BP independently measured with a laser-scanning microscope (see Supplementary).

Figure 3b shows the polarization dependence of the emission intensity measured at a constructive-interference position. The suspended BP film also exhibits anisotropic emission with a higher polarization contrast than that observed in the non-suspended thick BP region shown in Fig. 1b. The emission polarization is aligned parallel to the BP edge, indicating that the armchair crystal axis lies along the edge. Figure 3c shows the pump-fluence dependence of the spectrally integrated emission intensity under optical excitation polarized parallel to the armchair direction. The emission exhibits a pronounced nonlinear increase above a threshold pump fluence of approximately 0.8 mJ $cm^{-2}$ per pulse at 4 K. The threshold remains nearly constant up to 50 K but increases above 70 K. This nonlinear behavior is observable up to 150 K within the present pump fluence range. The corresponding emission spectra are shown in Fig. 3d. As the pump fluence approaches the threshold, a narrow and intense spectral peak appears at 4.65 μm. At higher pump fluences, the peak intensity increases further, and the spectral width broadens to shorter wavelengths, which is possibly caused by thermal drift during the spectrum acquisition. Figure 3e shows that the emission peak shifts toward shorter wavelengths as the temperature increases, which is consistent with the temperature dependence of the bandgap energy of BP [31]. Most importantly, time-resolved measurements provide more direct evidence of the underlying emission process. Figure 3f shows the emission dynamics measured below and above threshold of around 0.8 mJ $cm^{-2}$ per pulse. Below the threshold, the emission decays with a lifetime of 1.30 ns. However, above the threshold, an intense and ultrafast emission pulse appears immediately after optical excitation. As the pump fluence increases, the pulse width reaches approximately 110 ps, approaching the instrumental response limit. At the same time, a nanosecond-scale emission tail remains observable.

The spontaneous emission lifetime is nearly unchanged across the threshold. Instead, the emergence of intense picosecond emission pulses together with threshold behavior and spectral narrowing provides solid evidence for stimulated emission with positive optical gain. The increase of the threshold fluence above 70 K coincides with the temperature range where the transition of spontaneous emission from excitonic to EHP states occurs. If the optical loss is not changed with temperature, the observed increase in threshold indicates a reduction in optical gain at elevated temperatures. This behavior is consistent with the thermal dissociation of excitons and the subsequent disappearance of excitonic enhancement of the oscillator strength near the band edge. Such excitonic enhancement has previously been reported in the form of infrared absorption peaks in BP [32, 33]. In addition, the picosecond duration of the stimulated emission pulse also provides insight into the relevant carrier-loss mechanisms at that time. Since stimulated emission is completed within ~100 ps or less after excitation, optical gain is expected to be more sensitive to ultrafast Auger recombination than the slower nonradiative SRH recombination including surface recombination, whose characteristic timescale is close to 1 ns in the present samples [34]. Therefore, the relatively weak

Auger recombination predicted for BP helps to sustain the large population of photoexcited carriers required for positive optical gain at high pump fluences.

The observed stimulated emission can be assisted by optical feedback from a vertical Fabry-Pérot cavity formed in the suspended BP structure. Since the estimated cavity free spectral range (~0.88 μm) exceeds the spontaneous-emission bandwidth (~0.4 μm), the emitted MIR field is expected to couple predominantly to a single longitudinal cavity mode. (see Supplementary)

**Conclusion**

Our results establish that the MIR emission mechanism in BP evolves from excitonic recombination at cryogenic temperatures to EHP recombination at elevated temperatures. This crossover is reflected not only in the temperature dependence of the emission intensity but also directly in the emission dynamics. At low temperatures, the excitonic emission decay is governed by the nonradiative recombination mediated by surface states. Furthermore, the direct observation of intense picosecond stimulated emission demonstrates that BP can support ultrafast optical gain.

More broadly, this work demonstrates time-resolved upconversion spectroscopy as a powerful approach for investigating emission dynamics and related photoexcited carrier dynamics in narrow-bandgap semiconductors. Future studies using this method could provide further insight into many-body interaction under high density excitation, including Auger recombination and exciton-exciton interaction [35]. The ability to simultaneously resolve picosecond stimulated emission and nanosecond spontaneous emission processes contributes to the development of high-efficiency and high-speed MIR light emitting devices for sensing, communications, and integrated photonic technologies, which is inaccessible to conventional steady-state MIR spectroscopy.

**Methods**

Bulk BP flakes synthesized by HQ Graphene were mechanically exfoliated using adhesive tape and transferred onto a Si substrate coated with 90-nm-thick $SiO_2$. The BP flake was fixed with silver paste. After the transfer in ambient air, the sample was immediately mounted in the vacuum chamber of a liquid-helium flow cryostat.

The sample was excited with optical pulses from a mode-locked titanium-sapphire laser operating at a wavelength of 920 nm, a pulse width of 3 ps, and a repetition rate of 80 MHz. (see Supplementary) The pump light and the MIR emission from BP were separated using a long-pass dichroic mirror with a cut-off wavelength of 1800 nm. For high throughput microspectroscopy of small samples, we used a custom-made transmission objective lens with a magnification of 20 to focus the pump light and collect the MIR emission. The excitation spot diameter was estimated to be 3.6 μm with the knife-edge method. MIR emission spectra were acquired using Fourier-transform spectrometer equipped with a cooled indium antimonide (InSb) detector. The spectral resolution was 4.0 $cm^{-1}$, which corresponds to 8 nm in wavelength. To measure the spectrally integrated and polarization-dependent emission intensity, the emission was directly detected with another InSb detector. Time-resolved MIR measurements were performed using wavelength upconversion technique [36]. MIR photons emitted from BP were converted to near-infrared photons through intracavity sum-frequency generation in periodically poled lithium niobate pumped by a continuous-wave laser operating at 1064 nm (customized, NLIR ApS) [37]. The converted photons were detected by a superconducting single photon detector. The MIR detection range was 4.0 to 5.5 μm in wavelength. Emission transients were recorded using a time-correlated single-photon counting (TCSPC) unit triggered by the pump laser pulses. The background count was subtracted from the recorded data. The overall temporal resolution was 79 ps.

**Data availability**

All the data used in this work are available upon reasonable request.

**Acknowledgements**

We thank Dr. Masato Takiguchi for the development of the setup.

**Funding**

The authors declare no competing financial interest.

## FIGURES

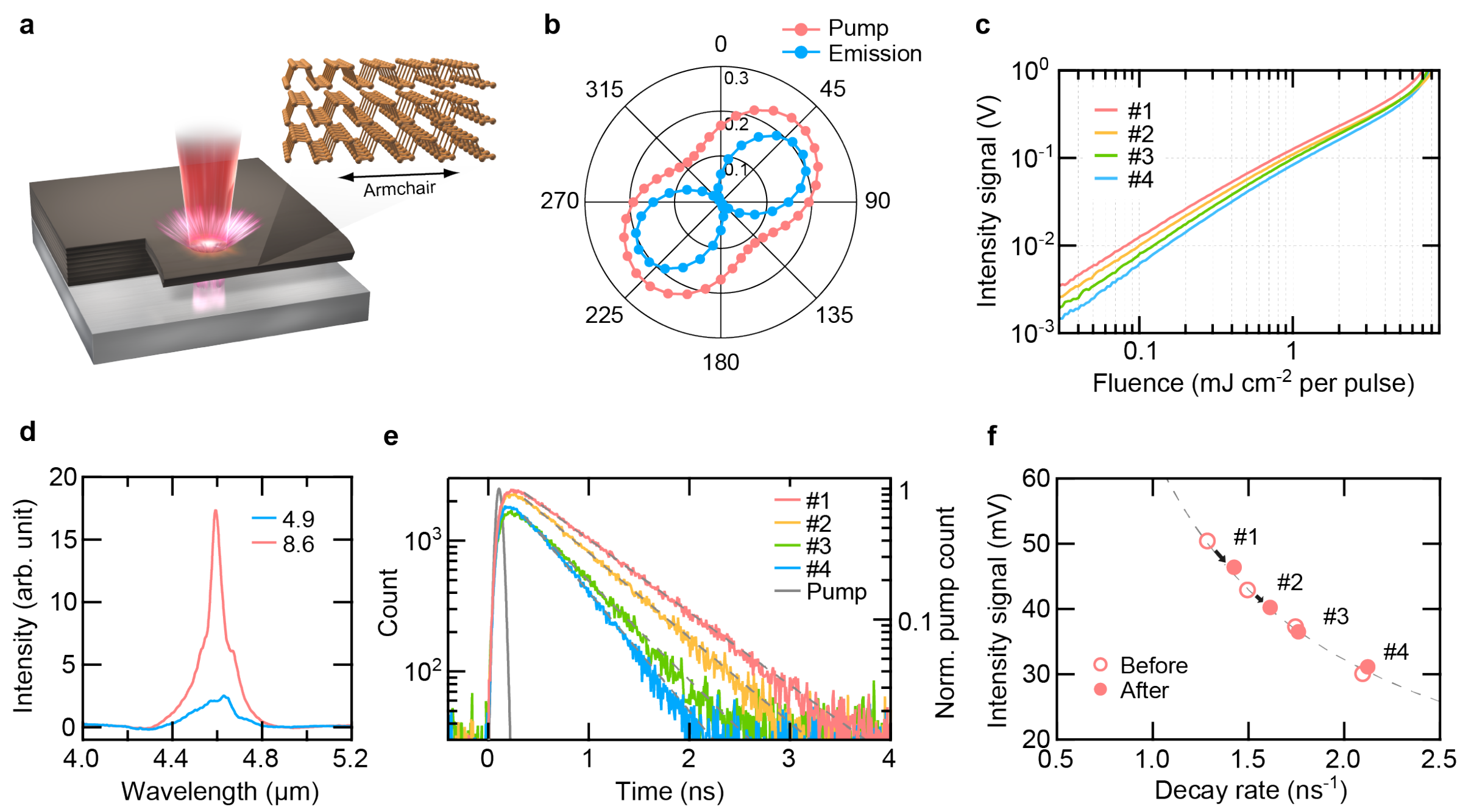

**Figure 1 | Spatial variation of emission properties in BP. a,** Schematic view of a BP flake. **b,** Polarization dependence of the MIR emission intensity measured at a pump fluence of 2.5 mJ cm$^{-2}$ per pulse. **c,** Spectrally integrated emission intensity measured at four different positions as a function of pump fluence. **d,** Emission spectra measured at pump fluences of 4.9 and 8.6 mJ cm$^{-2}$ per pulse. The spectral dip at 4.25 μm originates from atmospheric $CO_2$ absorption. **e,** Time-resolved emission measured at four different positions under excitation at 0.37 mJ cm$^{-2}$ per pulse. The dashed lines represent single-exponential fits. The full width at half maximum (FWHM) of the pump pulse is 79 ps. **f,** Spectrally integrated emission intensity as a function of emission decay rate before and after air exposure. During air exposure, the ambient temperature and humidity were 20 °C and 18%, respectively. The dashed curve represents a fit to the inverse relationship and is shown as a guide to the eye. All measurements were conducted at 4 K.

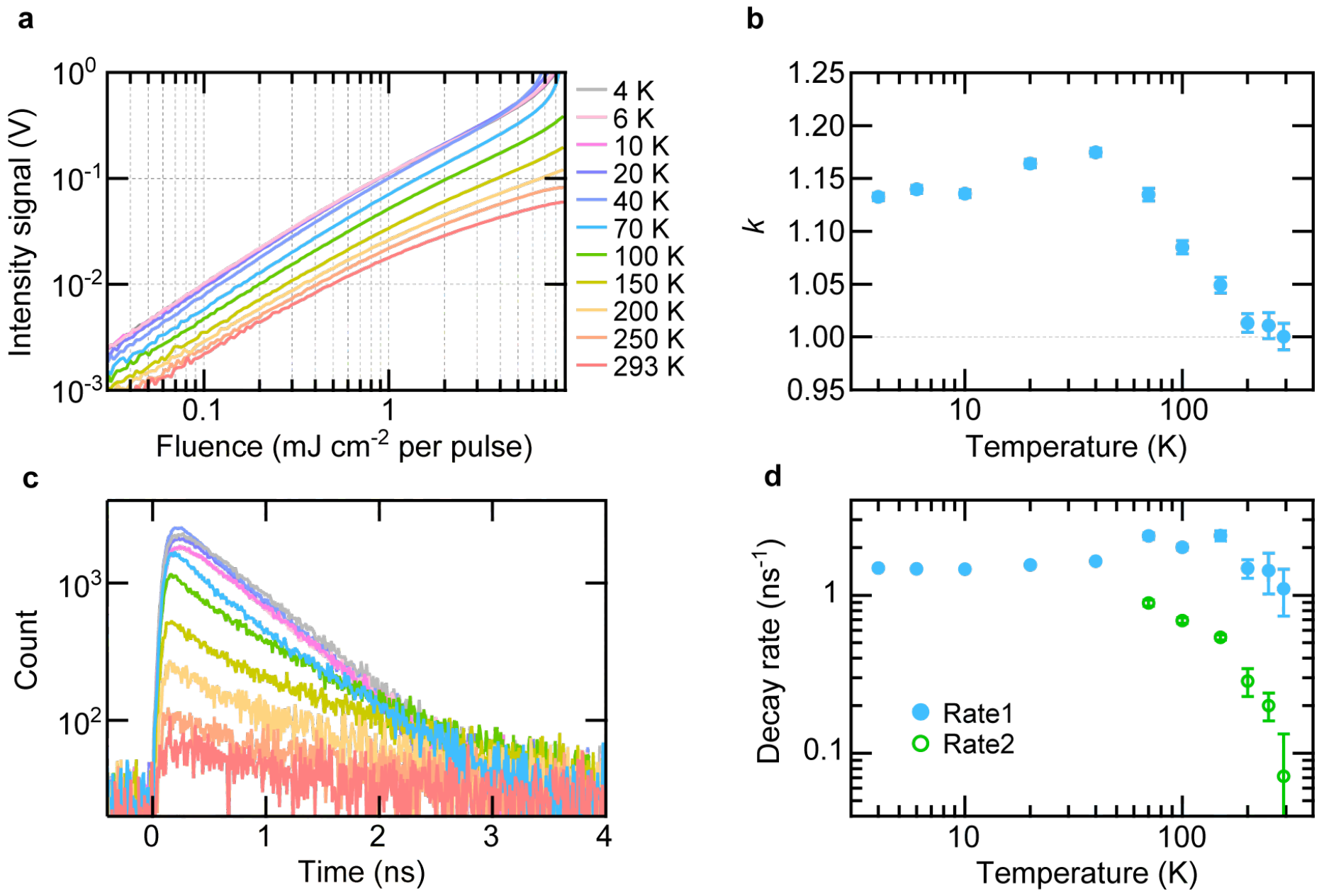

**Figure 2 | Temperature dependence of emission dynamics in BP. a,** Spectrally integrated emission intensity as a function of pump fluence at different temperatures. **b,** Power-law exponent $k$ obtained from the low-fluence regime (<0.2 mJ cm$^{-2}$ per pulse) in **a**. **c,** Time-resolved emission measured at different temperatures under excitation at 0.37 mJ cm$^{-2}$ per pulse. The line colors correspond to those in **a**. **d,** Fast and slow decay rate components extracted from the emission transients by single- or double-exponential fitting as a function of temperature. The error bars represent the uncertainties obtained from the curve fitting.

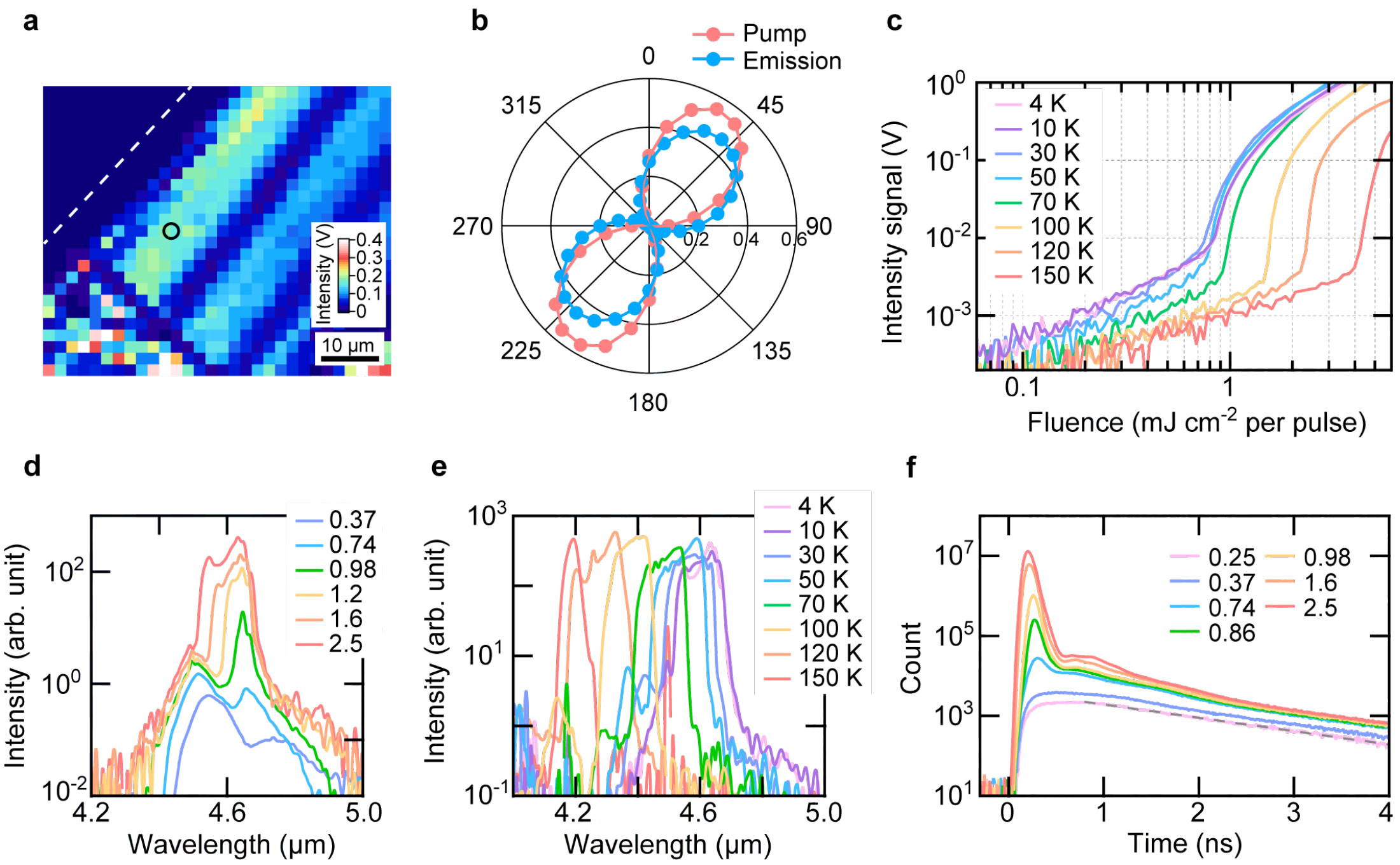

**Figure 3 | Pump-fluence and temperature dependence of emission properties in suspended BP. a,** Spatial map of the emission intensity near the edge of a BP flake measured at a pump fluence of 2.5 mJ cm$^{-2}$ per pulse and 4 K. The white dashed line and black circle indicate the edge of the BP flake and the measurement position, respectively. **b,** Polarization dependence of the emission intensity measured at a pump fluence of 2.5 mJ cm$^{-2}$ per pulse and 4 K. **c,** Spectrally integrated emission intensity as a function of pump fluence at different temperatures. **d,** Emission spectra measured at different pump fluences (mJ cm$^{-2}$ per pulse) at 4 K. **e,** Emission spectra measured at different temperatures under excitation at 2.5 mJ cm$^{-2}$ per pulse. **f,** Time-resolved emission measured at different pump fluences (mJ cm$^{-2}$ per pulse) at 4 K. The dashed line represents a single-exponential fit to the spontaneous emission regime.

# Mid-infrared spontaneous and stimulated emission dynamics in black phosphorus

**Supplementary Information**

## A. Estimation of photoexcited carrier density

The photoexcited carrier density is a key parameter for the quantitative analysis of carrier recombination dynamics. In the present microscopy setup, the excitation beam is tightly focused by an objective lens. Since black phosphorus (BP) exhibits fast and anisotropic carrier diffusion, the spatial distribution of photoexcited carriers expands rapidly within the instrumental temporal resolution, reducing the peak carrier density. The distribution of the focused excitation beam is assumed to be a two-dimensional Gaussian profile *f*,

$$f \propto \exp\left(-\frac{x^2+y^2}{2\sigma_0^2}\right), \tag{1}$$

where $x$ and $y$ denote the armchair and zigzag crystal directions, respectively. The standard deviation $\sigma_0$ is related to the full width at half maximum (FWHM) by

$$\sigma_0 = \mathrm{FWHM}_0 / 2\sqrt{2\ln 2}\,. \tag{2}$$

The carrier diffusion broadens the Gaussian distribution with time according to

$$\sigma_x = \sqrt{\sigma_0^2 + 2D_x t} \tag{3}$$

for the armchair direction, and

$$\sigma_y = \sqrt{\sigma_0^2 + 2D_y t} \tag{4}$$

for the zigzag direction. The corresponding FWHMs are given by

$$\mathrm{FWHM}_x = 2\sqrt{2\ln 2}\,\sigma_x\,, \tag{5}$$

$$\mathrm{FWHM}_y = 2\sqrt{2\ln 2}\,\sigma_y\,. \tag{6}$$

Within the first 80 ps after excitation, corresponding approximately to the instrumental temporal resolution, the circular carrier distribution evolves into an elliptical distribution owing to the anisotropic carrier diffusion in BP. The spot size of the excitation beam was measured to be $\mathrm{FWHM}_0$ = 3.6 μm with the knife edge method. Using the reported diffusion coefficients of $D_x = 1.3 \times 10^4$ and $D_y = 870$ cm$^2$/s for armchair and zigzag directions, respectively [1], the size of the broadened carrier distribution is estimated to be

$$\mathrm{FWHM}_x = 34\ \mu\mathrm{m},$$

$$\mathrm{FWHM}_y = 9.5\ \mu\mathrm{m}.$$

The area, $S = \pi \cdot \mathrm{FWHM}_x \cdot \mathrm{FWHM}_y/4$, is calculated to be $2.5 \times 10^{-6}$ $\mathrm{cm}^{-2}$.

The photoexcited carrier density $N(0)$ at the onset of the measured carrier recombination dynamics was estimated by using

$$N(0) = Abs \cdot P / (S \cdot \delta \cdot h\nu), \tag{7}$$

where $Abs$, $P$, $\delta$, and $h\nu$ are the optical absorption, pump energy per pulse (fluence), penetration depth, and photon energy of the pump light, respectively. The complex refractive index of BP along the armchair direction is reported to be $n = 3.6$ and $\kappa = 0.26$ at an excitation wavelength of 920 nm [2]. Using these values, the optical absorption and penetration length were estimated to be $Abs = 0.68$ and $\delta = 280$ nm, respectively. Here, neither carrier loss nor carrier multiplication is expected during the thermalization of photoexcited carriers [3].

**B. Estimation of radiative recombination coefficient**

At a sufficiently low photoexcited carrier density, Auger recombination can be neglected. In this regime, the photoexcited carrier density $N(t)$ is described by

$$dN/dt = - AN - BN^2. \tag{8}$$

Equation (8) has an analytical solution [4]

$$N(t) = N(0)e^{-At}/[1+ N(0)(B/A)(1-e^{-At})]. \tag{9}$$

Since the spontaneous emission intensity $I$ is proportional to $BN^2$,

$$I(t) = \zeta BN^2(t) \tag{10}$$

is obtained, where $\zeta$ is an emission extraction efficiency. The square root of the normalized emission intensity is described by

$$\sqrt{I(t)/I(0)} = N(t)/N(0). \tag{11}$$

Therefore, the curve fitting to the square root of the normalized emission transient with Eq. (9) enables the estimation of the radiative recombination coefficient $B$. Figure S1 shows the normalized time-resolved emission at 200 K. At a pump fluence of 0.37 mJ $\mathrm{cm}^{-2}$ per pulse, the carrier density at the onset of the measured recombination dynamics was estimated to be $N(0) = 1.7\times10^{18}$ $\mathrm{cm}^{-3}$. The curve fitting yields a radiative recombination of $B = (4.7 \pm 0.3) \times 10^{-10}$ $\mathrm{cm}^3\ \mathrm{s}^{-1}$. It was also found that $A$ is close to zero with a large uncertainty and could not be reliably determined by the curve fitting. This result indicates that the measured emission transient is dominated by two-particle (bimolecular) radiative recombination under the present experimental condition.

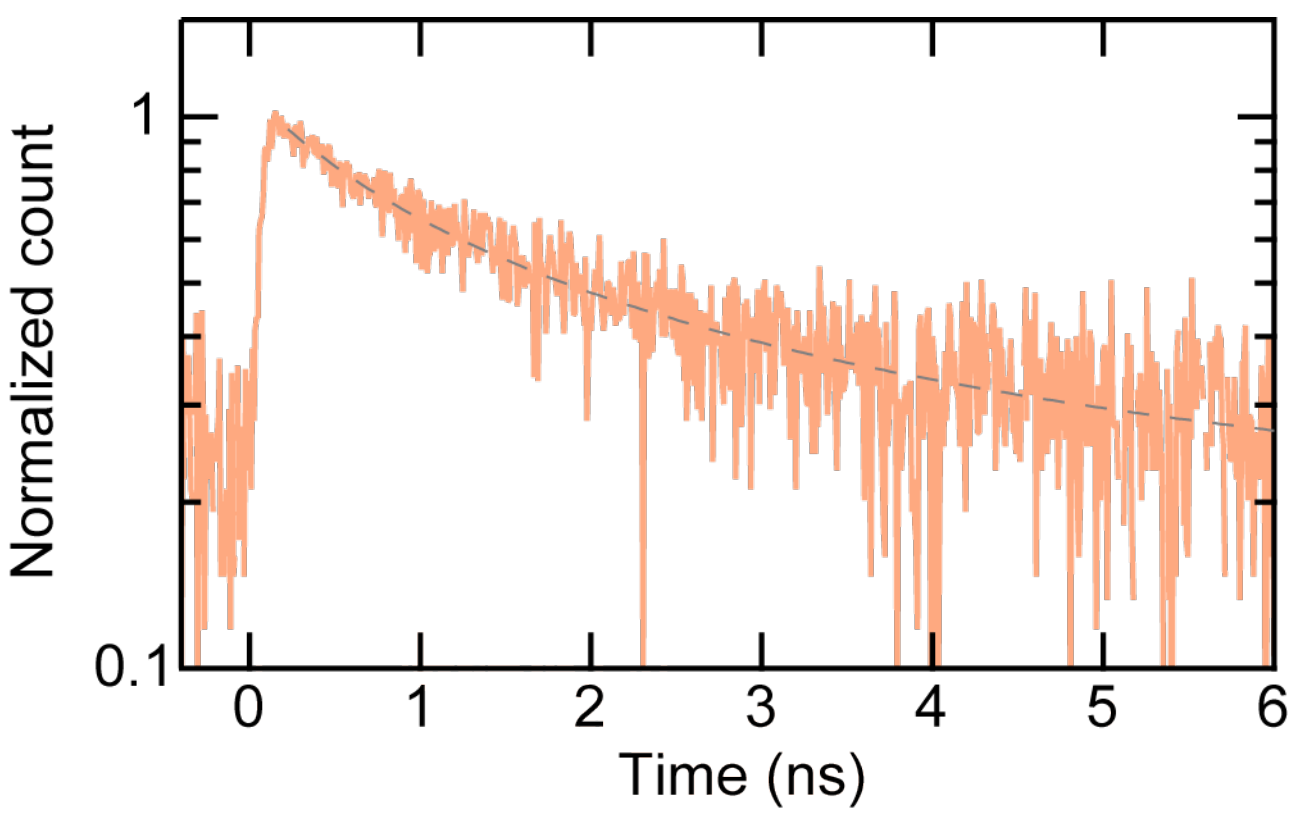

**Figure S1** | Normalized time-resolved emission measured at 0.37 mJ $cm^{-2}$ per pulse and 200 K. The dashed line represents a curve fit.

### C. Suspended BP structure

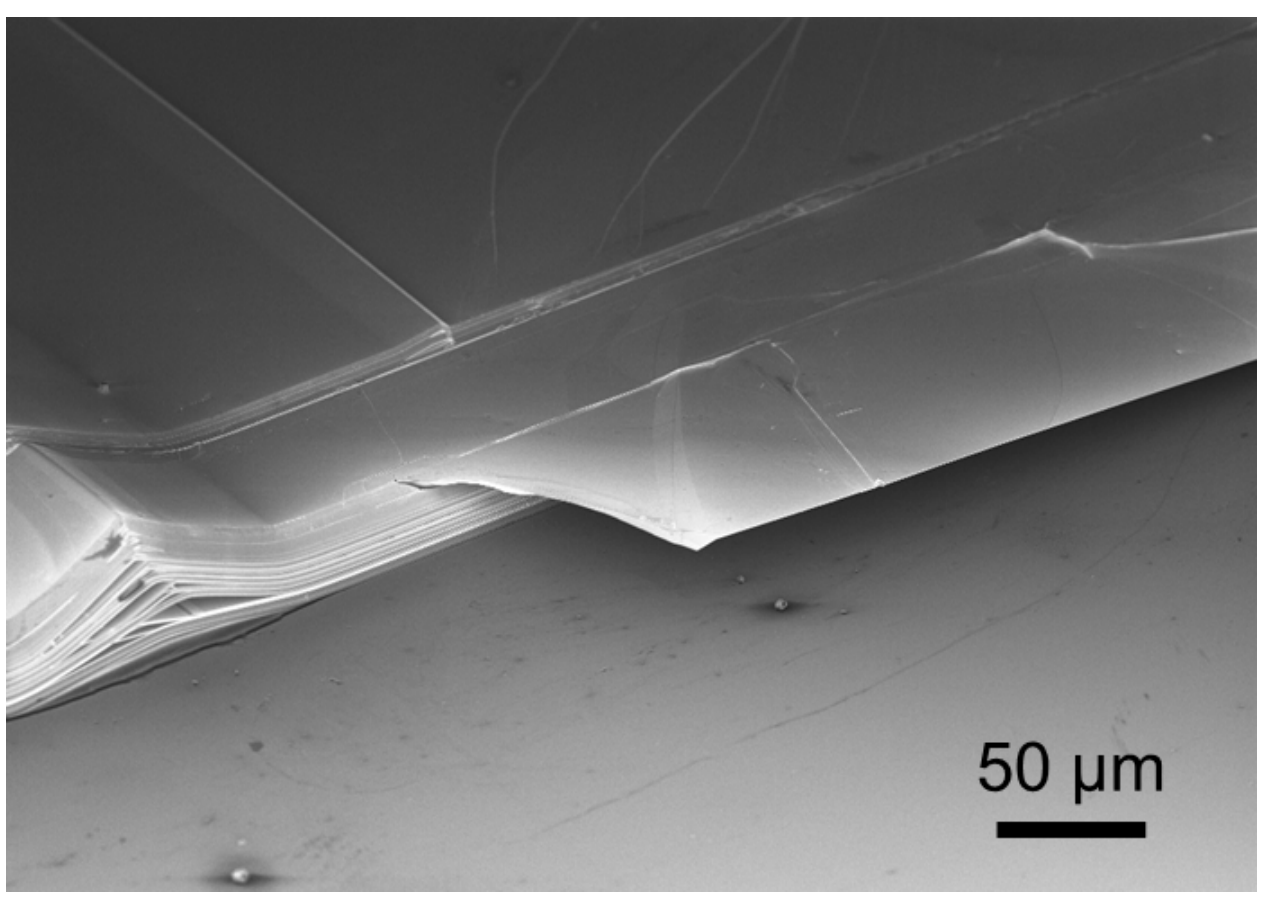

**Figure S2** | Scanning electron microscope image of the suspended BP structure in the vicinity of the measured position. The image was acquired at a tilt angle of 55°.

### D. MIR reflection spectrum

The suspended BP structure exhibits periodical peaks in the mid-infrared (MIR) reflection spectrum, as shown in Fig. S3. The spectral period is approximately 0.42 μm at a wavelength of 3.2 μm. Assuming a Fabry-Pérot cavity formed in the suspended BP structure, the cavity length ($L$) can be estimated from

$$L = \lambda^2/(2n\Delta\lambda), \qquad (12)$$

where $n$ is the real part of the refractive index of air, $\lambda$ and $\Delta\lambda$ are the wavelength and free spectral range (FSR), respectively. From Eq. (12) and $n$ = 1 for the air gap, the cavity length was estimated to be $L$ = 12 μm, which is in good agreement with the height of the suspended BP film measured by a laser-scanning microscope (Fig. S4). This agreement indicates that the observed interference of MIR emission at the suspended structure results from the formation of a vertical Fabry-Pérot cavity between the suspended BP film and the underlying $SiO_2$/Si substrate. Using the estimated cavity length, the FSR at a wavelength of 4.6 μm is calculated to be approximately 0.88 μm.

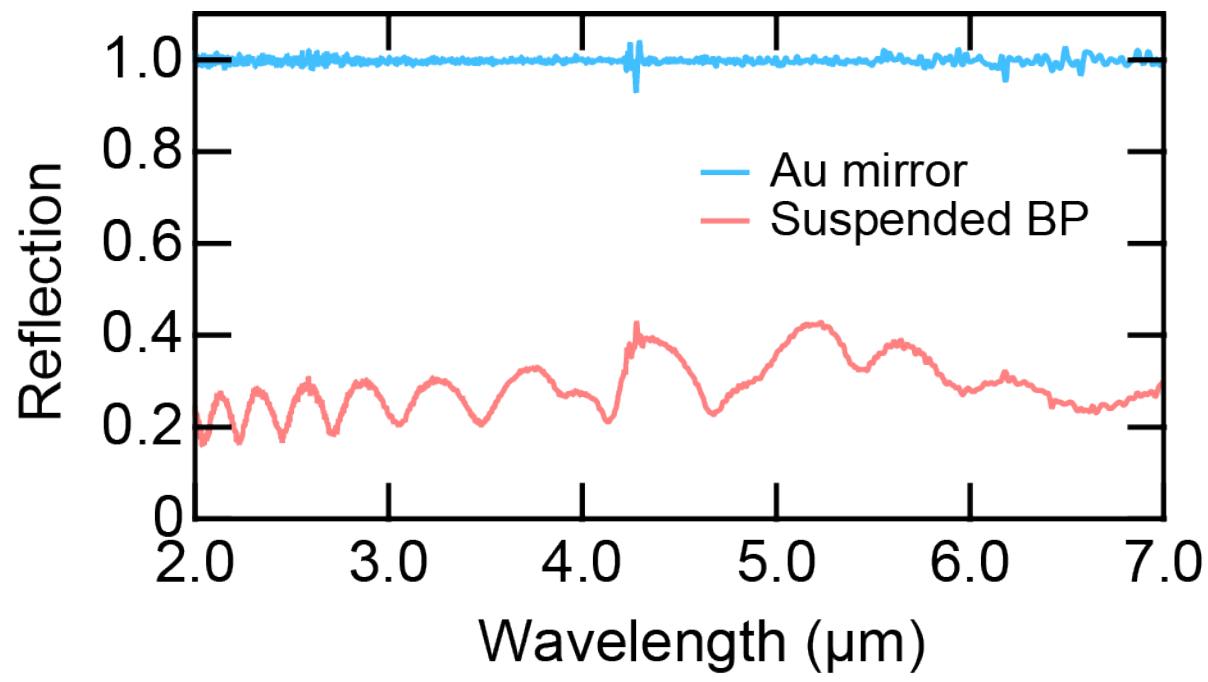

**Figure S3** | Mid-infrared reflection spectrum of the suspended BP structure at room temperature. The measurement area was defined by a 10 × 10 μm² aperture. The reflection spectrum of a gold-coated mirror is also shown as a reference. The small spectral spikes observed at 4.25 μm originate from atmospheric $CO_2$ absorption.

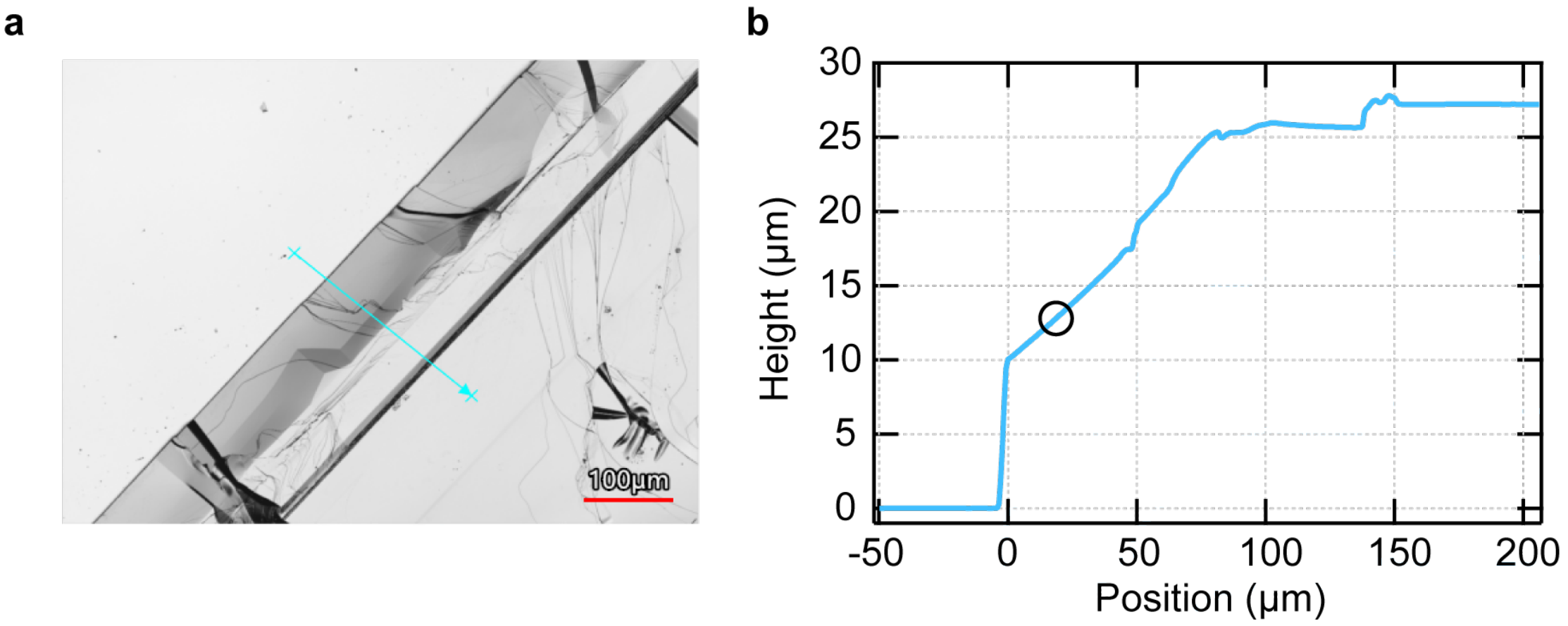

**Figure S4** | **a**, Laser-scanning microscope image of the suspended BP structure. The solid arrow indicates the measured position of the height profile. **b**, Height profile of the suspended BP structure. The position of MIR emission measurements is indicated by a solid circle.

## E. Setup

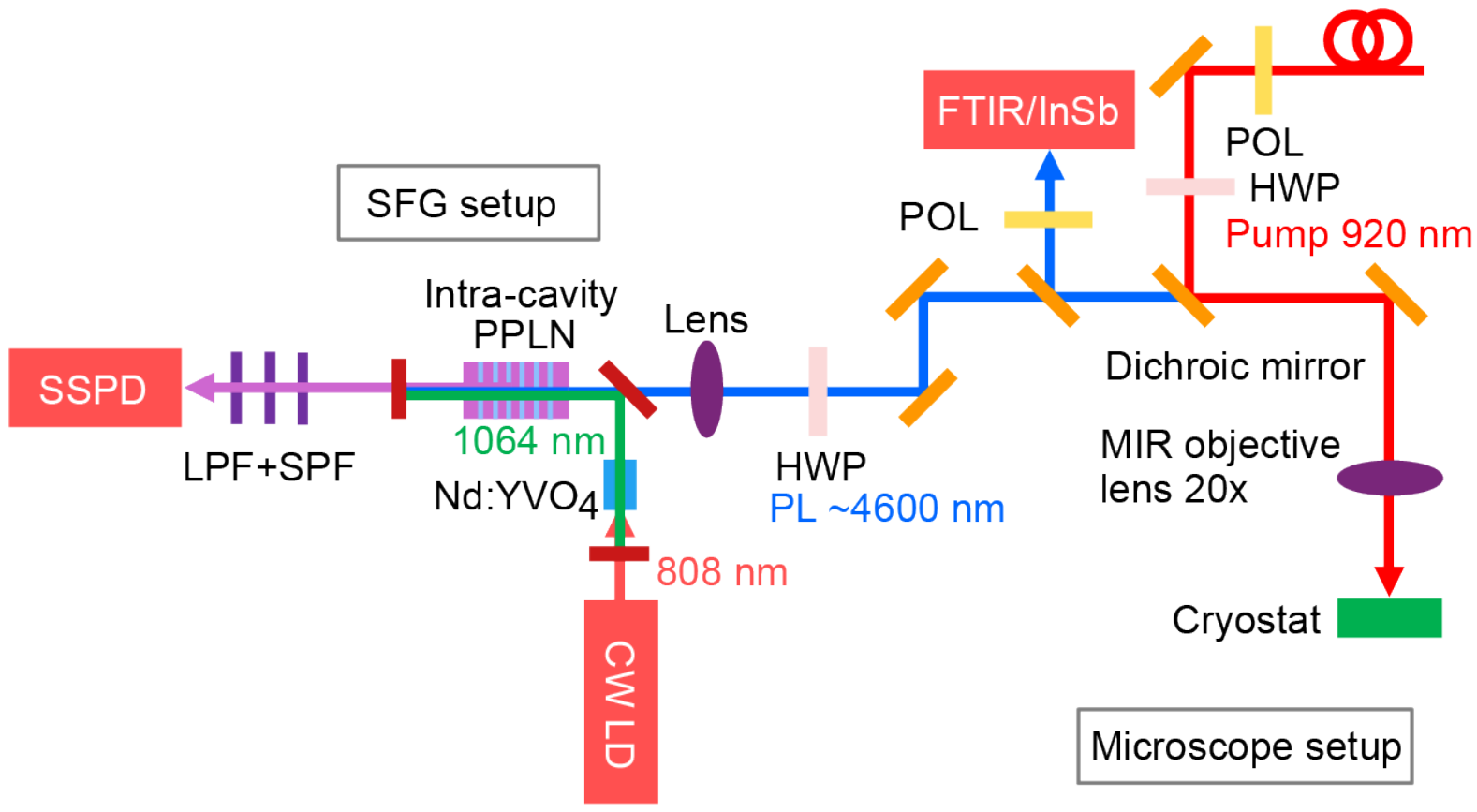

**Figure S5** | Schematic of the experimental setup. The setup consists of a microscopy unit and a sum-frequency-generation (SFG) unit. SSPD denotes a superconducting single photon detector. LPF and SPF denote long-pass and short-pass optical filters, respectively. HWP and POL denote half-wave plates and polarizers, respectively.